\documentclass[%
 aps ,prl,twocolumn, superscriptaddress, nobalancelastpage
]{revtex4-1}

\usepackage{graphicx}
\usepackage{dcolumn}
\usepackage{bm}

\usepackage[utf8]{inputenc}
\usepackage[T1]{fontenc}
\usepackage{mathptmx}
\usepackage{etoolbox}
\usepackage{subfigure}
\usepackage{appendix}
\usepackage{comment}
\usepackage{amsmath}
\usepackage[colorlinks=true,citecolor=blue]{hyperref}

\makeatletter
\def\@email#1#2{%
 \endgroup
 \patchcmd{\titleblock@produce}
  {\frontmatter@RRAPformat}
  {\frontmatter@RRAPformat{\produce@RRAP{*#1\href{mailto:#2}{#2}}}\frontmatter@RRAPformat}
  {}{}
}%
\makeatother

\begin{document}

\preprint{AIP/123-QED}
\title{Quantum Dot Source-Drain Transport Response at Microwave Frequencies}
\author{Harald Havir}
\address{NanoLund and Solid State Physics, Lund University, Box 118, 22100 Lund, Sweden}
\author{Subhomoy Haldar}
\address{NanoLund and Solid State Physics, Lund University, Box 118, 22100 Lund, Sweden}
\author{Waqar Khan}
\address{NanoLund and Solid State Physics, Lund University, Box 118, 22100 Lund, Sweden}
\author{Sebastian Lehmann}
\address{NanoLund and Solid State Physics, Lund University, Box 118, 22100 Lund, Sweden}
\author{Kimberly A. Dick}
\address{NanoLund and Solid State Physics, Lund University, Box 118, 22100 Lund, Sweden}
 \address{Center for Analysis and Synthesis, Lund University, Box 124, 22100 Lund, Sweden}
\author{Claes Thelander}
\address{NanoLund and Solid State Physics, Lund University, Box 118, 22100 Lund, Sweden}
\author{Peter Samuelsson}
\address{Physics Department and NanoLund, Lund Universityd, Box 118, 22100 Lund, Sweden}
\author{Ville F. Maisi}
\address{NanoLund and Solid State Physics, Lund University, Box 118, 22100 Lund, Sweden}

\email{harald.havir@ftf.lth.se}
\date{\today}

\begin{abstract}
 Quantum dots are frequently used as charge sensitive devices in low temperature experiments to probe electric charge in mesoscopic conductors where the current running through the quantum dot is modulated by the nearby charge environment.
 Recent experiments have been operating these detectors using reflectometry measurements up to GHz frequencies rather than probing the low frequency current through the dot.
 In this work, we use an on-chip coplanar waveguide resonator to measure the source-drain transport response of two quantum dots at a frequency of 6 GHz, further increasing the bandwidth limit for charge detection. 
 Similar to the low frequency domain, the response is here predominantly dissipative. For large tunnel coupling, the response is still governed by the low frequency conductance, in line with Landauer-Büttiker theory. 
 For smaller couplings, our devices showcase two regimes where the high frequency response deviates from the low frequency limit and Landauer-Büttiker theory: When the photon energy exceeds the quantum dot resonance linewidth, degeneracy dependent plateaus emerge. 
 These are reproduced by sequential tunneling calculations. In the other case with large asymmetry in the tunnel couplings, the high frequency response is two orders of magnitude larger than the low frequency conductance $G$, favoring the high frequency readout. 
\end{abstract}

\maketitle

\begin{figure*}
    \centering
    \includegraphics[width = \textwidth]{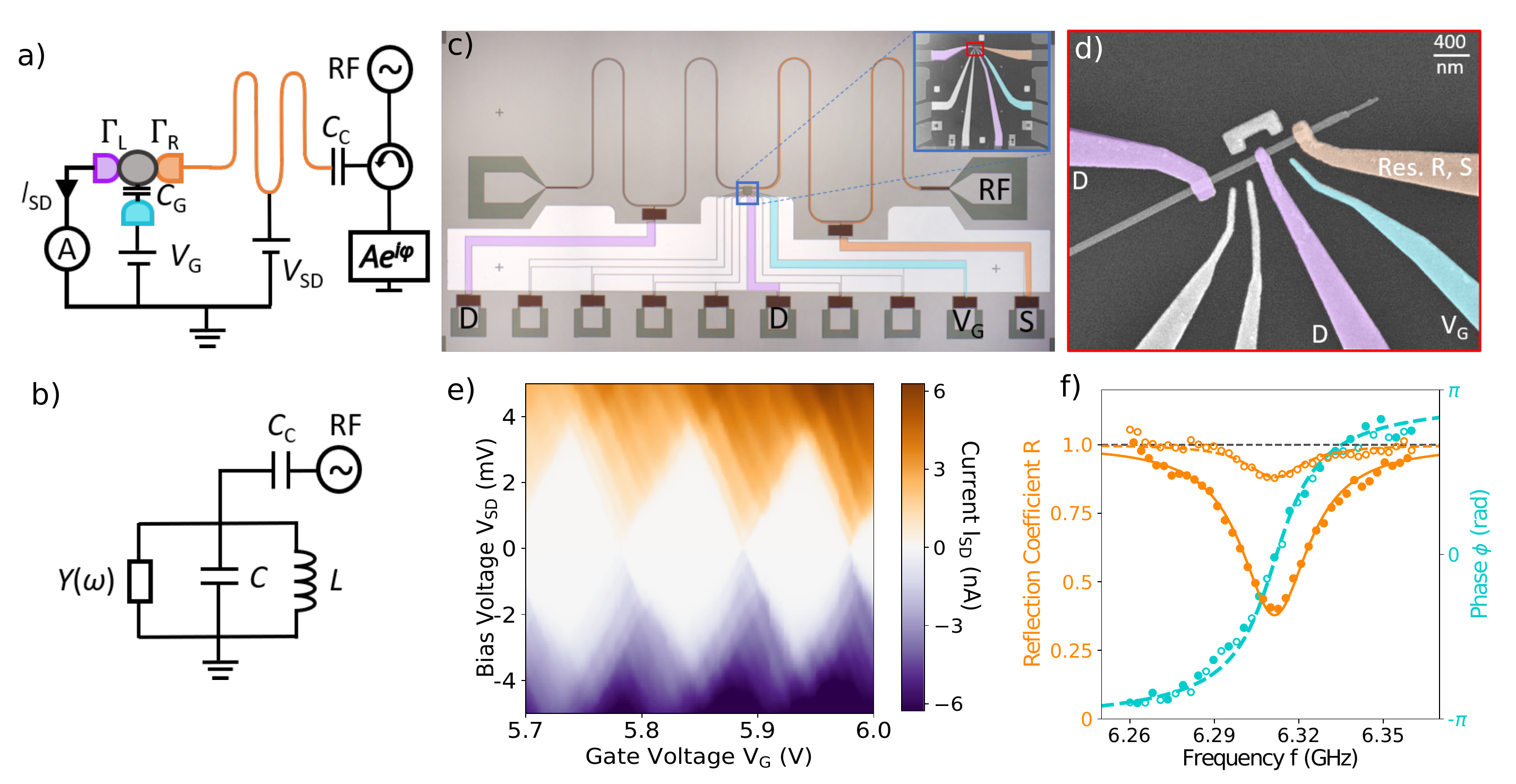}
    \caption{
    \textbf{(a)} A schematic diagram of the studied device. A microwave resonator (orange) is driven with a high-frequency signal (RF) through the coupling capacitor $C_C$ and the reflected amplitude $A$ and phase $\phi$ is measured. A QD with tunnel couplings $\Gamma_R$ and $\Gamma_L$, connects to the resonator via the source contact. The DC electrical current $I_{SD}$ is measured from the drain contact and DC voltage bias $V_{SD}$ is applied via the resonator and a gate voltage $V_G$ via a separate gate electrode.
    \textbf{(b)} The equivalent lumped-element LC circuit for the device with the complex admittance $Y(\omega)$ arising from the QD.
    \textbf{(c)} An optical micrograph of the device. The microwave resonator and DC lines are defined using a Nb etch-back method. The DC lines are capacitively shunted towards the resonator with a 30 nm aluminium oxide - 50 nm aluminium stack (white area). The contacts nearby the QD, visible in the scanning electron micrograph of the inset, are defined using EBL and deposited using Ni/Au evaporation.
    \textbf{(d)} A zoom-in of panel c showing the InAs nanowire in which the QD is defined.
    \textbf{(e)} The measured detector current $I_\mathrm{SD}$ as a function of bias and gate voltages $V_\mathrm{SD}$ and $V_\mathrm{G}$.
    \textbf{(f)} Measured reflection coefficient $R$ and phase $\phi$ as a function of frequency $f$ with the QD in Coulomb blockade (open circles) and conducting at zero bias voltage (dots) at $V_\mathrm{G} = 5.9$ V. 
    Solid lines are fits to Eq. (\ref{eq:Reflection_Coefficient}) with $f_r = $ 6.315 GHz, $\kappa_\mathrm{C}/2\pi = $ 22.3 MHz and $\kappa_\mathrm{QD} = 0$ for Coulomb blockade and $\kappa_\mathrm{QD}/2\pi = 4.7$ MHz for the QD in resonance.
    } 
    \label{fig:Figure1}
\end{figure*}

\section{Introduction}
The ability to detect single electrons in the solid state is useful for a variety of applications, including spin qubit readout \cite{Kane1998, SpinQubitsQDs, Vandersypen2017, Hanson2007}, electrical current and capacitance standards \cite{CapacitanceStandard, RedefiningAmpereReview}, studying cooper pair breaking \cite{vanWoerkom2015, Mannila2022, Ranni2021}, single-shot photodetection \cite{Gustavsson2007, Ghirri2020, Khan2021, Cornia2023}, and nanothermodynamics and fluctuations~\cite{Koski2014, Koski2014b, Barker2022, CountingInBLG, Kung2012, Manzano2021}. 
While many methods exist to detect charge, one of the main ways are by utilizing quantum dots (QD).
These systems make excellent charge detectors due to their high sensitivity and well-established transport theory \cite{IhnBook2009,Davies_book}, allowing detectors to be made predictable and with a well-understood operation principle.
Originally, measurements were performed at DC, relying on a difference in current for the readout resulting in a bandwidth up to some kHz \cite{RedefiningAmpereReview, Lafarge1993}.
In the last two decades, the readout methods have moved towards measuring the reflected power in a high-frequency tank circuit with resonant frequency in the 100 MHz - 1 GHz range.
This results in bandwidths in the MHz range allowing for $\mathrm{\mu}$s time resolution \cite{RFSET_Shoelkopf, LowFrQDReflectometry1, LowFrQDReflectometry2}.
The response of the system in these studies is still governed by the low frequency response of the system, i.e. the admittance $Y(\omega)$ is equal to the DC conductance $G$ of the system. 
In this article, we increase the QD sensor frequency to the 4 - 8 GHz frequency range where the cavity photon energy $\hbar\omega$ is greater than the thermal energy $kT$ ~\cite{Li2018}.
This opens up the avenue to increase the bandwidth correspondingly by an order of magnitude, yielding possibly a time resolution sufficient to probe the electron position in DQD systems within the recently achieved coherence times~\cite{Mi_Petta_Decoherence2016, StocklausserDecoherence2017}.
The pioneering works have considered the dispersive response of the QD at these frequencies motivated mostly by quantum capacitance effects \cite{FreqyQD_resonator}.
In this article, we focus on the dissipative part that yields a stronger response, making it useful for charge readout~\cite{Li2018}.
We present experimental results for two devices and show that for both of them at sufficiently large tunnel couplings that we are lifetime broadened, $\Gamma > kT$, the low frequency result of $Y(\omega) = G$ still applies.
However, when the device is tuned to the thermally broadened limit where the tunnel couplings $\Gamma < kT$, the measured admittance is qualitatively different from the DC conductance, displaying a linewidth of $2\hbar\omega$ in the QD level tuning and a factor two difference in admittance depending on the direction of the level shift of the quantum dot relative to the leads $\varepsilon$, attributed to spin degeneracy.
These results are well captured by sequential tunneling theory, directly evaluating the admittance for a QD subjected to a time-periodic drive \cite{BruderAndShoeller}, or using $P(E)$ theory in which the admittance is inferred from the absorption in the cavity \cite{Ingold1992, Souquet2014}.
Lastly, we show in the other device which exhibits asymmetric tunnel couplings where the DC transport is suppressed while remaining lifetime broadened, the AC response in this device remains large, in line with Ref.~\citealp{Li2018}, indicating a potentially useful consequence of probing QD devices at high frequencies.
This response falls in a regime where neither non-interacting scattering theory nor sequential tunneling models are applicable.

\section{Device Configuration}
The main device used to perform measurements is illustrated schematically in Fig \ref{fig:Figure1} a). 
The device builds on a transmission line resonator, shown in orange, which for the fundamental mode is equivalent to the LC circuit of panel b).
The right end of the resonator is connected to an input  line via a coupling capacitor $C_\mathrm{C}$, which allows the measurement of the amplitude and phase of a reflected signal.
The left end of the resonator on the other hand couples to a QD via the right junction capacitance.
This configuration makes the QD source-drain transport admittance $Y(\omega)$ appear directly on the LC resonator. 
At low drive frequency $\omega = 2\pi f$, this admittance is given just by the DC conductance $G$, i.e. $Y(\omega) = G$, and the QD gives rise to dissipation in the resonator.

The reflection coefficient of the input port is given by (see Appendix A)
\begin{equation}
\label{eq:Reflection_Coefficient}
    R = 1 - \frac{(\kappa_{\mathrm{QD}}+\kappa_\mathrm{i})\kappa_\mathrm{C}}{\left(\kappa/2\right)^2 + \left(\omega - \omega_\mathrm{r} -\delta\omega_\mathrm{\; QD}\right)^2},
\end{equation}
where $\omega_\mathrm{r} = 1/\sqrt{LC}$ is the resonance frequency, 
$\kappa = \kappa_{\mathrm{QD}} + \kappa_\mathrm{i} + \kappa_\mathrm{C}$ the sum of all the couplings defining the linewidth of the resonance, 
$\kappa_\mathrm{i}$ the internal losses, 
$\kappa_\mathrm{C} =Z_0\omega^2C_\mathrm{C}^2/C$ the input coupling \cite{goeppl} and 
$\kappa_\mathrm{QD} = \mathrm{Re} (Y(\omega))/C$ the QD coupling strength. 
The term $\kappa_\mathrm{QD}$ is directly proportional to the admittance $\mathrm{Re} (Y(\omega))$, determining $\kappa_\mathrm{QD}$ from a change in the measured reflection coefficient $R$ thus allows us to determine the dissipative part of the QD response $\mathrm{Re}(Y(\omega))$.
On the other hand, $\mathrm{Im}(Y(\omega))$ gives rise to a disperse shift $\delta\omega_\mathrm{\; QD} = \mathrm{Im}(Y(\omega))/2C$ of Eq.~(\ref{eq:Reflection_Coefficient}), which results in a change in the resonance frequency. These dispersive shifts are typically small, of the order of $10^{-3}\omega_r$ as is also the case for our devices, and have been studied in detail for QDs coupled capacitively via a gate electrode \cite{FreqyQD_resonator}. 

To measure the DC conductance $G$ of the QD at the same operation point as $Y(\omega$), we apply a DC bias voltage $V_{\mathrm{SD}}$ to the voltage node point in the middle of the $\lambda/2$ resonator such that it does not disturb the resonance, but appears at the source contact of the QD~\cite{FreqyQD_resonator, Khan2021}.
The current $I_{\mathrm{SD}}$, measured from the drain contact, yields then the conductance $G = \mathrm{d}I_{\mathrm{SD}}/\mathrm{d}V_{\mathrm{SD}}$ and enables the comparison of this low frequency transport result to the high frequency admittance $Y(\omega)$.
These DC lines, in addition to a gate line with applied gate voltage $V_\mathrm{G}$ to change the electron number in the QD, are shunted with a large capacitor to ground to prevent microwaves leaking out from the lines.

The physical realization of the device is presented in Fig. \ref{fig:Figure1} c). 
The coplanar waveguide, highlighted in orange, is a 9.86 mm long metallic strip of width 10 $\mathrm{\mu}$m with a gap of 5 $\mathrm{\mu}$m to the ground plane. 
Based on Ref. \citealp{goeppl}, we estimate the lumped-element capacitance $C = 765$ fF and inductance $L = 871$ pH, giving a characteristic impedance $Z_0 = \pi/2\sqrt{L/C} = 53$ $\mathrm{\Omega}$ \cite{goeppl}.
An RF port connects to the resonator with a $400\ \mu$m long two-finger geometry which defines the input coupling $\kappa_\mathrm{C}$.
The QD forms in an epitaxially grown InAs nanowire, see Fig \ref{fig:Figure1} d), by altering the growth between zincblende (ZB) and wurzite (WZ) crystal phase~\cite{NWGrowth}.
The WZ segments have a conduction band offset of $135$ meV compared to the ZB segments \cite{BandOffset}, forming tunnel barriers and a ZB segment between the barriers defines the QD with length 130 nm and diameter of 80 nm \cite{QDinNW}.
The location of these barriers is discerned by selectively growing GaSb on the ZB segments which highlights the features of the QD \cite{DavidB}.
The offset between ZB and WZ allow the atomically sharp definition of barriers leading to a well-defined QD. 
The DC lines are capacitively shunted by growing a 30 nm thick aluminium oxide layer with atomic layer deposition and evaporating 50 nm thick aluminium film to the light-gray area in Fig \ref{fig:Figure1} c).
Additional inductive filtering is added to all the DC pads as well as the midpoint connections to reduce RF leakage \cite{petta_filtering}.
The device is bonded to a printed circuit board and measured in a dilution refrigerator at the electronic temperature of $T= $ 50 mK at base temperature. 
Figure \ref{fig:Figure1} e) shows the measured Coulomb diamonds exhibiting a charging energy $E_\mathrm{C} = 3.5$ meV and excited states with energies around $300$ $\mathrm{\mu}$eV.
The lever arm to the gate, $\alpha = 0.03$ eV/V, is also determined. 

Figure \ref{fig:Figure1} f) presents the resonator response with the QD in Coulomb blockade (CB) and in the conduction resonance at $V_{G} = 5.9$~V which is attained with an input power $P = $ -130 dBm to the resonator.
With the QD transport suppressed in CB (open symbols), we determine the bare resonator properties by fitting these data to Eq.~(\ref{eq:Reflection_Coefficient}) with $\kappa_\mathrm{QD} = 0$ and $\delta\omega_\mathrm{QD} = 0$.
We obtain the resonance frequency $f_r = 6.318$ GHz, as well as determine the coupling strengths $\kappa_C/2\pi = 22.3$~MHz, and $\kappa_i/2\pi = 0.7$~MHz. 
The phase response (cyan rings) shows a $2\pi$ winding, characteristic for an overcoupled resonator.
Next, the QD is tuned to resonance  and the measurements are repeated. 
The corresponding data (solid markers) demonstrate that the linewidth of the resonance increases due to additional absorption in the QD.
The reduction in amplitude also reflects the increase of total dissipation in the resonator by $\kappa_\mathrm{QD}$.
Keeping the resonator parameters acquired from the previous data set fixed, the fit to Eq.~(\ref{eq:Reflection_Coefficient}) is now repeated to provide $\kappa_{QD}/2\pi = 4.7$ MHz and vanishing $\delta\omega_\mathrm{QD}$.

\section{Comparison of conductance and high frequency admittance}

\begin{figure}
    \centering
    \includegraphics[width = 0.5\textwidth]{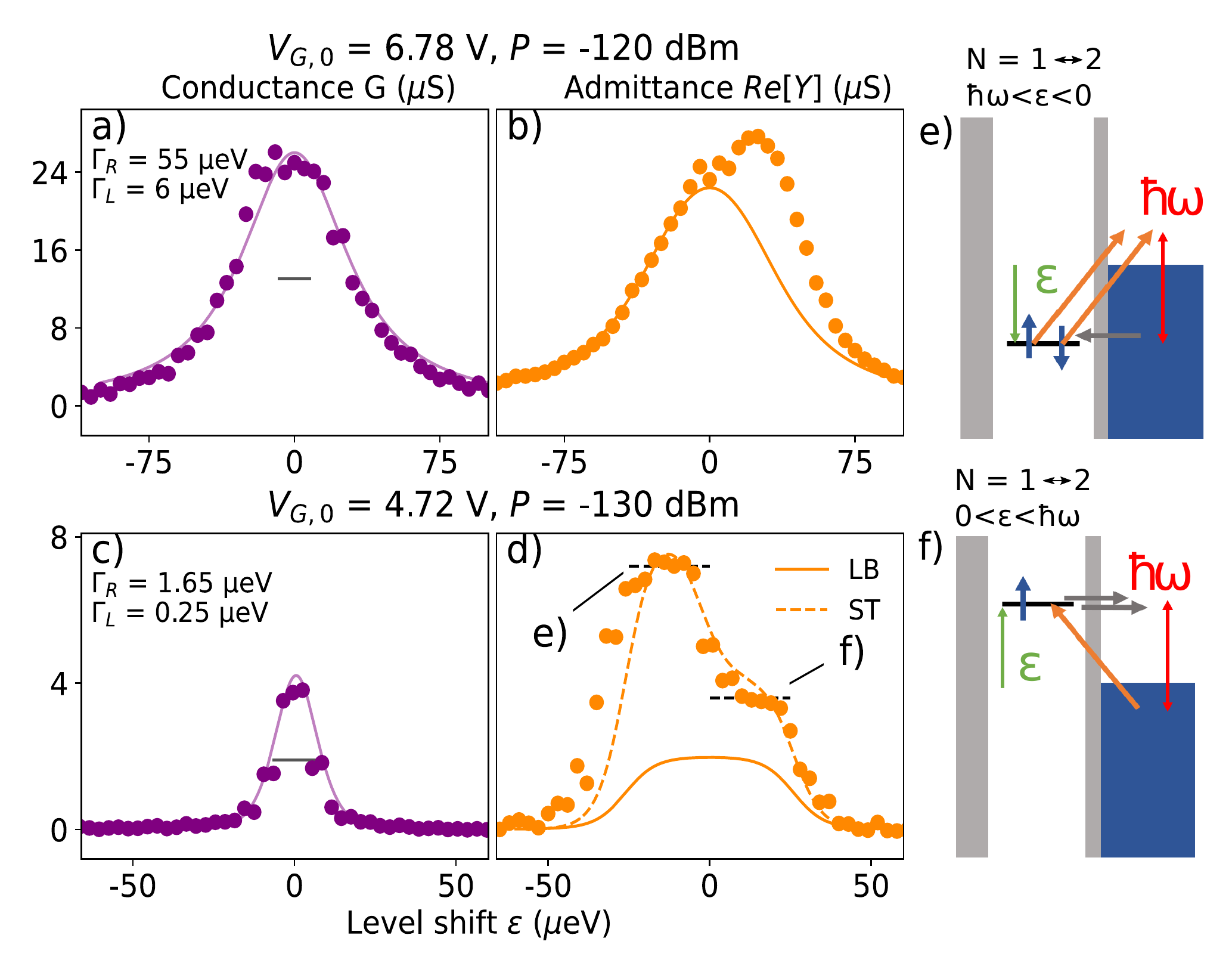}
    \caption{\textbf{(a)} The DC conductance $G$ measured at zero bias ($V_\mathrm{SD}=0$) as a function of the level shift $\epsilon = -\alpha(V_\mathrm{G}-V_\mathrm{G0})$ for the QD resonance at $V_\mathrm{G0} = $6.78 V without applying the RF drive. A small gate voltage $V_{\mathrm{G},2} = 0.5$ V is applied to the two remaining gate contacts in order to tune the tunnel couplings of the single dot slightly. The line shows a fit to Landauer-Büttiker (LB) theory, Eq. (\ref{DC_Lifetime_fit}). 
    \textbf{(b)} The admittance data $\mathrm{Re}(Y(\omega))$ around the same resonance as in a). The solid line is the finite-frequency Landauer B\"uttiker theory of Eq. (\ref{GR}). 
    \textbf{(c, d)} Data for another resonance at $V_\mathrm{G0} = 4.72$ V. The fit in panel c) is done using Eq. (\ref{eq:diffCondBS}), and in panel d) the dashed line is a Sequential Tunneling (ST) calculation of Eq.~(\ref{eq:BnS_ST_theory_equation}) based on the formalism in Ref.~\citealp{BruderAndShoeller}.
    \textbf{(e, f)} Band diagrams with the two energy level $\varepsilon$ settings corresponding to the two plateaus of panel d). The orange arrows indicate tunneling processes involving photon absorption, while the grey arrows indicate the tunneling events returning the system back to the lowest energy state shown with the blue arrows. The number of orange/grey arrows specify the number of electrons which can participate in the corresponding tunneling process between the $N=1$ and $N=2$ electrons on the dot.}
    \label{fig:SymQD}
\end{figure}

Now we turn to comparing the low frequency conductance $G$ and the high-frequency response $Y(\omega)$ presented in Figs. \ref{fig:SymQD} a) and b) for a lifetime-broadened resonance at $V_\mathrm{G}=$ 6.7 V.
The QD conductance $G$ has a peak width of $60$ $\mu$eV $>$ 4 kT, hence, a fit (solid line) to Landauer B\"uttiker theory~\cite{Davies_book} yields the tunnel couplings $\Gamma_\mathrm{L}=$ 6 $\mathrm{\mu}$eV and $\Gamma_\mathrm{R} =$ 55 $\mathrm{\mu}$eV. 
In Fig. \ref{fig:SymQD} b) the measured admittance $\mathrm{Re}(Y(\omega)) = \kappa_{QD}C$ is shown for the same resonance with an input power $P=$ -120 dBm.
This admittance response is identical to the DC conductance $G$ within 30 $\%$, as expected from the low-frequency prediction of $Y(\omega) \approx G$.
A numerical calculation based on Landauer-Büttiker theory (orange line) for this system, see appendix B and Ref.~\cite{Prtre1996DynamicAO}, also predicts the equivalence $\mathrm{Re}(Y(\omega)) = G$ for this configuration.

The energy of a single microwave photon is $\hbar\omega_\mathrm{r} = 26$ $\mu$eV, hence in the configuration of Fig. \ref{fig:SymQD} a)-b) the lifetime broadening exceeds the photon energy.
By reducing the gate voltage, the tunnel coupling reduces, decreasing the lifetime broadening.
This allows us to make the linewidth of the DC conductance peak smaller than the photon energy.
Tuning from $V_G = 6.78$ V to $V_G = 4.72$ V results in a thermally broadened peak, shown in Fig. \ref{fig:SymQD} c).
The Landauer B\"uttiker theory fit again reproduces the results with $\Gamma_\mathrm{L} = $ 0.25 $\mu$eV.
As the linewidth is now set by temperature and not the tunnel couplings, the larger coupling may vary from $\Gamma_\mathrm{R} = $ 0.5 to 6 $\mu$eV without disrupting the fit to the data.
Now the measured $Y(\omega)$, presented in Fig. \ref{fig:SymQD} d), shows a broader peak with a qualitatively different peak shape than the conductance has, hence, the equivalence $Y(\omega) = G$ is broken.
The response has two plateaus at 2.3 and 4.6 $\mu$S, extending out by 30 $\mu$eV to either direction from the midpoint, and matching the photon energy in line with the DC current response studied in Ref.~\citealp{PhotoassistedTransport1994}.
The broadening of $Y(\omega)$ arises since with the energy of the photon, the system overcomes an additional charging energy cost up to $\hbar\omega_r$ as depicted in Figs. \ref{fig:SymQD} e) and f).
The factor of two difference arises from the spin degeneracy in the QD. 
The photon absorption rate is twice for tunneling out of the QD (applies for $\epsilon >0$) as compared to tunneling into the QD  (applies for $\epsilon < 0$). 
A sequential tunneling model with either time dependent voltage drive~\cite{BruderAndShoeller} or P(E) theory~\cite{Ingold1992,Souquet2014} describes the full response.
The two theories agree at low resonator - QD couplings, but as the coupling increases (e.g. by increasing cavity impedance), the P(E) theory predicts spontaneous emission events which further change the transport.
See Appendices C and D for details.
Here we have fitted the value of the larger tunneling rate to $\Gamma_R = 1.65\ \rm{\mu eV}$ which sets the overall height of the response.
The spin degeneracy shifts the resonance point and the total height of the resonance peak by a small amount, see Eq. \ref{eq:p}, thus the DC fitting parameter values were adjusted to $\Gamma_L = 0.25\ \mu$eV and $\alpha\Delta V_{G0} = -1.5\ \mu$eV.

\begin{figure}
\centering
\includegraphics[width = 0.5\textwidth]{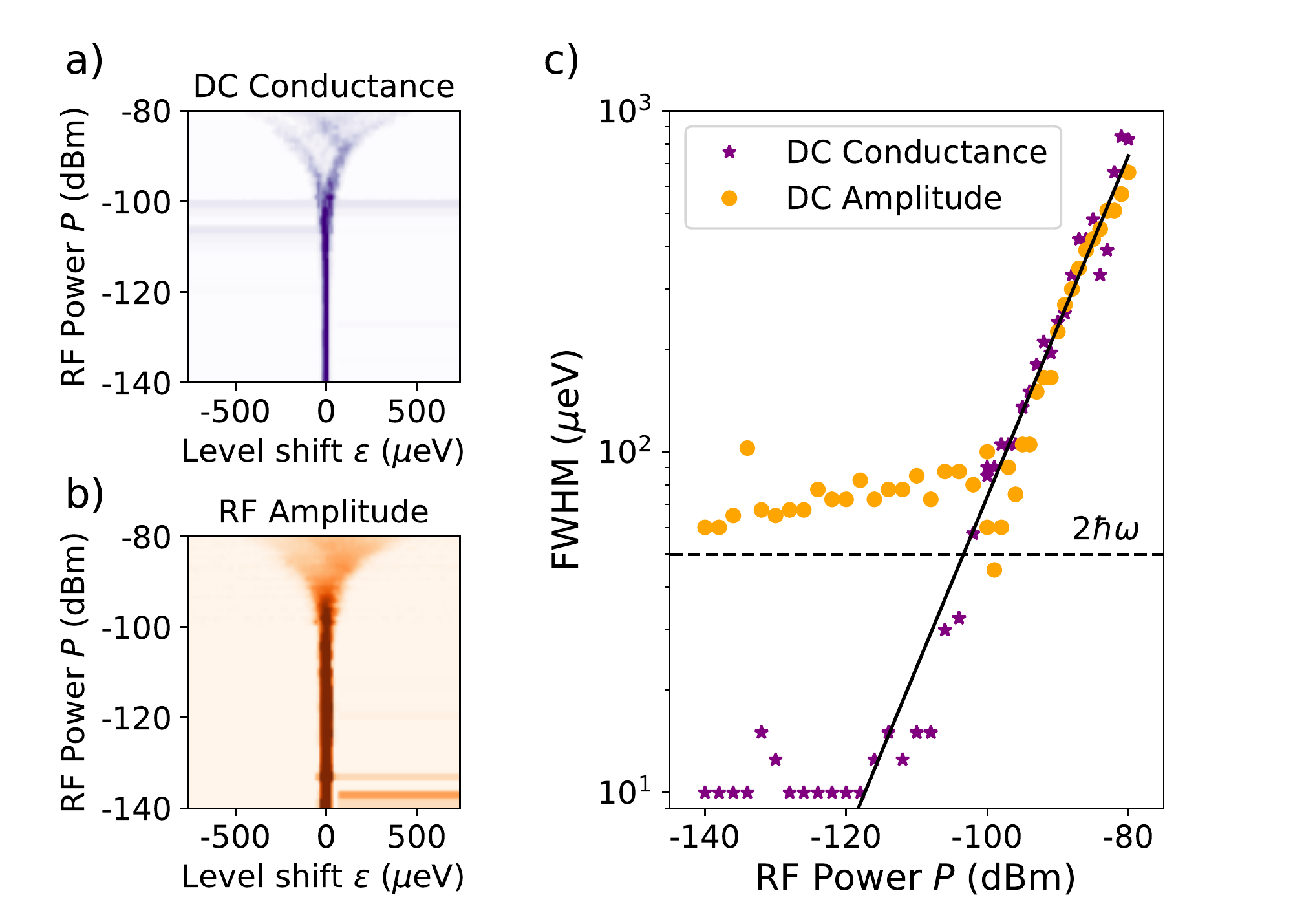}
\caption{The DC conductance \textbf{(a)} and RF amplitude \textbf{(b)} are measured simultaneously as the microwave power to the resonator input is changed. The measured linewidths are plotted in \textbf{(c)} along with a dashed black line indicating $2\hbar\omega_r$ and a solid black line corresponding to the calculated microwave amplitude of Eq.~(\ref{eq:Vamp}).}
\label{fig:PowerPlot}
\end{figure}

Figure \ref{fig:PowerPlot} extends the measurements of Figs. \ref{fig:SymQD} c) and d) as a function of drive power $P$.
At low drives $P < -100$ dBm, the linewidth of $Y(\omega)$ is essentially set by $2\hbar\omega_\mathrm{r}$ while the linewidth of $G$ remains thermally broadened for $P < -120$ dBm.
At high power additional broadening is observed in both the DC and RF result, in line with the previous works by Refs.~\citealp{modena_amplitude, Frey_Power}. 
The amplitude of the microwave oscillations inside the resonator is estimated by following the steps of Refs.~\citealp{Amplitude_Equation, Haldar2023} yielding the microwave amplitude
\begin{equation}
\label{eq:Vamp}
V_\mathrm{MW} = \left(\frac{4Q^2Z_0}{Q_\mathrm{ext}}P \right)^{-1/2}.
\end{equation}
With characteristic impedance $Z_0 = \pi/2\sqrt{L/C} = 53$ $\Omega$, quality factor $Q = \omega_r/\kappa = 262$ and external quality factor $Q_\mathrm{ext} = 270$, we obtain the solid line in Fig. \ref{fig:PowerPlot} c).
As the voltage amplitude $V_\mathrm{MW}$ enters the high power regime, the broadening of the RF and DC response both arise from the amplitude of the microwave signal.
For the microwave response, the boundary point $P \approx -100 $ dBm between the high and low power regime is set by the condition $eV_\mathrm{MW} = \hbar\omega_r$, i.e. whether the energy related to the amplitude or single photon is dominant.
However, the width of the DC feature continues to be defined by the drive amplitude until the power $P = -120 $ dBm, at which point the energy corresponding to the drive amplitude becomes smaller than the thermal energy i.e. $eV_{MW} < kT$.
The measurements of $Y(\omega)$ performed in the measurements of Fig. \ref{fig:SymQD} were performed at $P \leq -120$ dBm, allowing high-power effects to be ignored in the analysis.
Note also that the DC measurements of Fig. \ref{fig:SymQD} were performed without applied microwave drive, though applying the drive does not change the DC response at this power level.

\begin{figure}
    \centering
    \includegraphics[width = 0.5\textwidth]{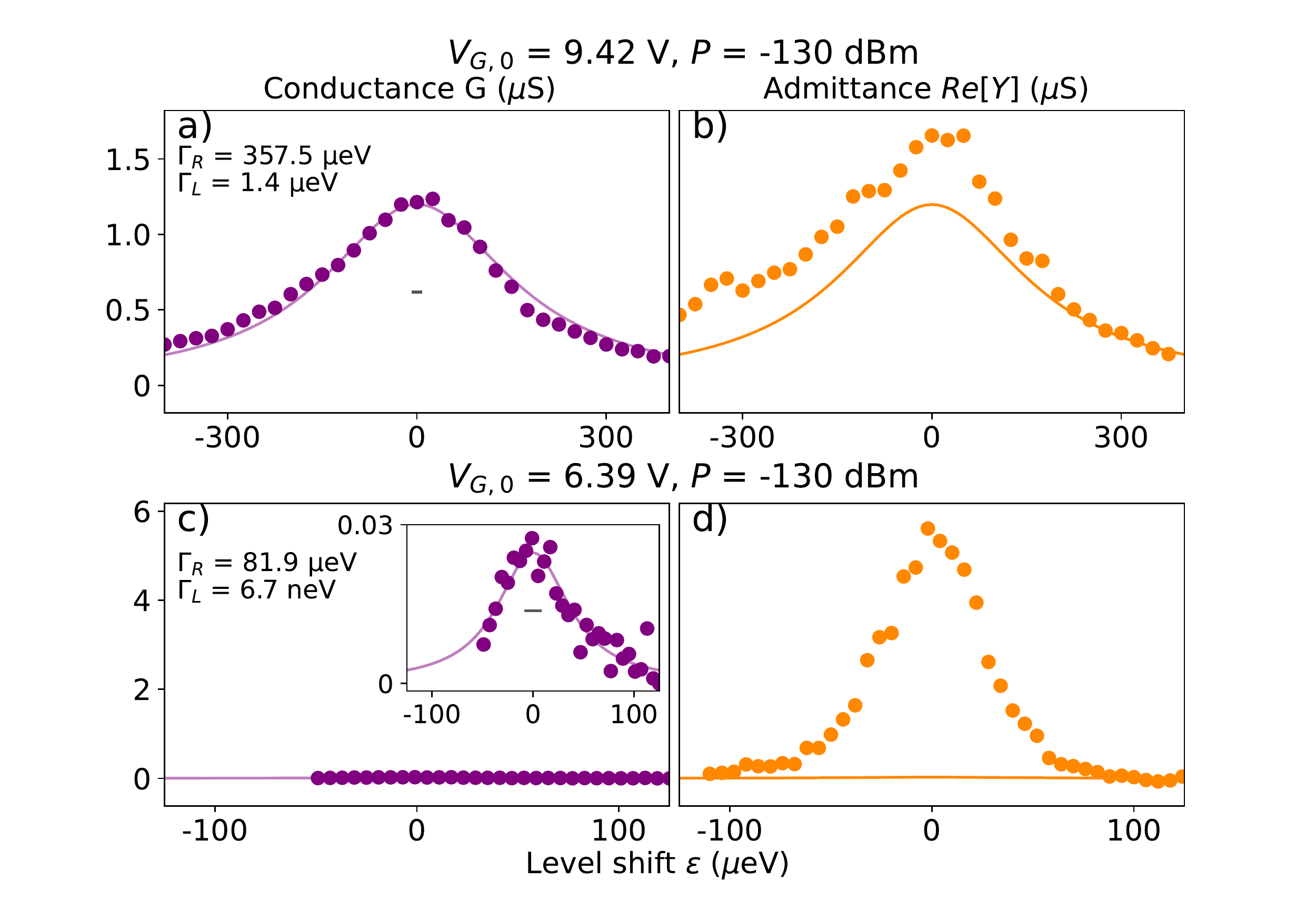}
    \caption{The same measurements as of Fig. \ref{fig:SymQD} for a second device at $V_{\mathrm{G}0} = 9.4$~V in \textbf{(a)} and \textbf{(b)}, and $V_\mathrm{G0} = 6.4$~V in \textbf{(c)} and \textbf{(d)}. For this second QD, the charging energy and lever arm are $E_C = 2$~meV, and $\alpha = 0.04$ eV/V. }
    \label{fig:AsymQD}
\end{figure}

Figures \ref{fig:AsymQD} a) and b) repeat the study in the lifetime broadened case for a second device.
Now we have a much more asymmetric device with fitted values of $\Gamma_\mathrm{L} = 1.4\ \mu$eV and $\Gamma_\mathrm{R} = 357.5\ \mu$eV catching again the equivalence $Y(\omega) = G$, valid in both experiment and theory.
Tuning the QD to a lower gate voltage $V_{\mathrm{G}_0} = 6.73$ V, Figs. \ref{fig:AsymQD} c) and d), has again the effect of reducing the tunnel barriers such that the conductance $G$ is suppressed by two orders of mangitude. This results in a correspondingly smaller $\Gamma_\mathrm{L} = 6.7$ neV while the right barrier $\Gamma_\mathrm{R} = 80$ $\mu$eV still provides a lifetime broadening to the system.
In the measured high-frequency response, we observe a peak with the same linewidth as the DC feature but with an amplitude value of $Y(\omega)|_{\epsilon = 0} = 3.6$ $\mu$S, which is two orders of magnitude greater than the peak value of the DC conductance of $30$ nS.
In this case, Landauer-Büttiker theory still predicts $Y(\omega) = G$. Therefore the linewidth of $Y(\omega)$ of fig \ref{fig:AsymQD} d) is reproduced correctly but the predicted overall magnitude is two orders of magnitude lower than the measured response. 
The sequential tunneling calculations miss lifetime broadening effects, thus not replicating the linewidth. 
The predicted peak admittance of $Y(\omega) = 30$ $\mu S$ on the other hand predicts qualitatively correct that the RF response is stronger, though the predicted value is an order of magnitude larger than the measured. With these arguments and findings, we interpret that the correct picture to describe is closer to the sequential tunneling case where the microwave drive is divided between the junction capacitances and then a considerable fraction of the drive arises across the transparent junctions and leads to large dissipation as observed before for lower frequencies~\cite{Chorley2012}. The Landauer-Büttiker theory differs from this as the total admittance of the system determine the voltage division, which in our case would lead to the same dissipation as at DC. To describe the response quantitatively, a more advance theory combining the above aspects would be needed~\cite{Ridley2022}.

\section{Conclusions}
In summary, we studied the high frequency source-drain response of a quantum dot. We showed experimentally that the low frequency result of $Y(\omega) = G$ holds 
for quantum dots tuned to sufficiently large tunnel couplings in line with the slow-drive limit. However, when the tunnel couplings are tuned to be smaller than the photon energy, 
the measured linewidth of the admittance $Y(\omega)$ is set by the photon energy.
This response is well-described by sequential tunneling theory.
Additionally, the low-frequency limit does not hold when the drive amplitude is made sufficiently large or with large asymmetry in tunnel couplings of the junctions.
For the highly asymmetric case, it is also shown that the admittance $Y(\omega)$ can be orders of magnitude larger than the conductance $G$, indicating a potential benefit of measuring at high frequencies, as the readout strength remains large even for weakly conducting dots.

\section{Acknowledgements}
We thank for financial support from the Foundational Questions Institute, a donor advised fund of Silicon Valley Community Foundation (grant number FQXi-IAF19-07), the Knut and Alice Wallenberg Foundation through the Wallenberg Center for Quantum Technology (WACQT), Swedish Research Council (Dnr 2019-04111) and NanoLund.

\bibliography{bibliography.bib}


\onecolumngrid
\setcounter{equation}{0}
\appendix*
\renewcommand{\theequation}{A.\arabic{equation}}

\section{Appendix A: Lumped element circuit and microwave reflection probability}
\label{Appendix A}
The QD-resonator system, linearly driven at a frequency $\omega$ close to resonance, is modelled as a lumped element circuit, shown in Fig. \ref{fig:Figure1} b). The resonator is described by an inductance $L$ and capacitance $C$ and internal losses are accounted for by a resistance $R_i$. The QD has a frequency dependent complex admittance $Y(\omega)$. The resonator is coupled to an input transmisison line, with impedance $Z_0$, via a coupling capacitance $C_\text C$. The total impedance $Z(\omega)$ of the circuit is then given by
\begin{equation}
Z=\frac{1}{i\omega C_\text C}+\left(Y(\omega)+1/R_i+i\omega C +\frac{1}{i\omega L}\right)^{-1}\equiv Z_\text R+i Z_\text I.
\end{equation} 
Writing $\omega_r=1/\sqrt{LC}, \kappa_\text i=R_i/C, \kappa_{\text{QD}}=\text{Re}[Y(\omega)]/C$, and $\delta \omega_{\text{QD}}=\text{Im}[Y(\omega)]/(2C)$ and using that $\omega \approx \omega_r$ we can write the real and imaginary parts of the impedance as
\begin{eqnarray}
Z_\text R&=&\frac{(\kappa_{\text{QD}}+\kappa_i)/(4C)}{(\kappa_{\text{QD}}+\kappa_\text i)^2/4+(\omega-\omega_\text r+\delta \omega_\text{QD})^2} \\
Z_\text I&=&-\left(\frac{1}{\omega_\text{r} C_{\text C}}+\frac{(\omega-\omega_\text r+\delta \omega_\text{QD})/C}{(\kappa_{\text{QD}}+\kappa_\text i)^2/4+(\omega-\omega_\text r+\delta \omega_\text{QD})^2}\right). \nonumber
\end{eqnarray} 
The reflection probability $R$ for a coherent microwave drive tone at $\omega$ is given by 
\begin{equation}
R=\left| \frac{Z(\omega)-Z_0}{Z(\omega)-Z_0}\right|^2 =1-\frac{4Z_RZ_0}{(Z_R+Z_0)^2+Z_\text I^2}
\label{refl}
\end{equation}
and the corresponding reflection phase is
\begin{equation}
\phi=\mbox{atan}\left(\frac{2Z_\text IZ_0}{Z_R^2+Z_\text I^2-Z_0^2}\right)
\end{equation}
For the reflection probability, inserting $Z_{\text R}$ and $Z_{\text I}$ into Eq. (\ref{refl}), writing the capacitive couping rate $\kappa_{\text C}=Z_0\omega_r^2C_\text{C}^2/C$ and neglecting terms proportional to the small parameter $C_\text{C}\omega_rZ_0 \ll 1$, we arrive at
\begin{equation}
R=1-\frac{(\kappa_{\text{QD}}+\kappa_\text{i})\kappa_\text{C}}{(\kappa/2)^2+(\omega-\omega_\text{r}^*+\delta \omega_\text{QD})^2},
\end{equation}
where we introduced the total $\kappa=\kappa_{\text{QD}}+\kappa_\text{i}+\kappa_\text{C}$ and $\omega_\text{r}^*=\omega_r(1+C_\text{C}/C)$, the capacitive coupling renormalized resonance frequency. By noticing that $C_\text{C}/C\ll1$ we can put $\omega_\text{r}^*\approx \omega_\text{r}$ and we then arrive at Eq. (\ref{eq:Reflection_Coefficient}) in the main text.

\section{Appendix B: Landauer-B\"uttiker Formalism}
\label{Appendix B}
\begin{figure}[h]
  \centering
  {\includegraphics[scale=0.5]{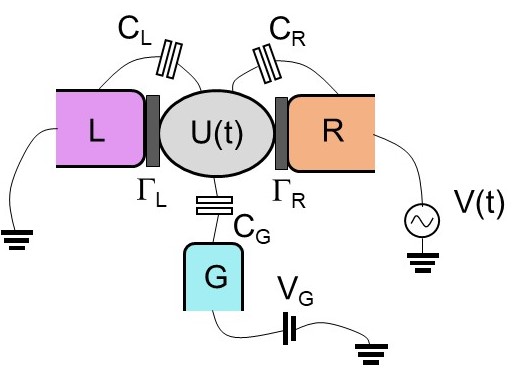}}
  \caption{Sketch of the quantum dot showing, in addition to Fig. \ref{fig:Figure1} of the main article, the tunnel junction capacitances $C_\text{L},C_\text{R}$ and the applied potential $V(t)$ on the right contact and induced potential $U(t)$ in the QD.}
\label{syspic}
\end{figure}

An extended sketch of the main article Fig. \ref{fig:Figure1} a) is presented in Fig. \ref{syspic}.  It includes junction capacitances and potentials.
In this appendix, we calculate the QD admittance within the Landauer-Büttiker formalism. 
With this approach, the QD admittance $Y(\omega)$ is evaluated within a time-dependent scattering approach, neglecting Coulomb blockade effects but fully accounting for the current conservation at the QD via the flow of dynamic screening currents. Our result is an extension of the discussion presented by Pretre, Thomas and B\"uttiker, \cite{Prtre1996DynamicAO}, here including QD-lead capacitive couplings. We therefore present only the main steps in the derivation. 

The staring point for the calculation is the energy dependent, symmetric scattering matrix $S(E)$ of the QD, assuming effectively a single transport channel, given by
\begin{equation}
S(E)=\left(\begin{array}{cc} r(E) & t'(E) \\ t(E) & r'(E) \end{array}\right),
\end{equation}
where the reflection and transmission amplitudes are given by the Breit-Wigner expressions 
\begin{eqnarray}
r(E)&=&1-\frac{i\Gamma_L}{E-\epsilon+i(\Gamma_L+\Gamma_R)/2} \nonumber \\ 
r'(E)&=&1-\frac{i\Gamma_R}{E-\epsilon+i(\Gamma_L+\Gamma_R)/2} \nonumber \\
 t(E)&=&t'(E)=\frac{i\sqrt{\Gamma_L\Gamma_R}}{E-\epsilon+i (\Gamma_L+\Gamma_R)/2}.
\label{smat}
\end{eqnarray}
Here $\epsilon=\epsilon_\text{d}-\alpha V_\text{G}$ is the energy of the discrete QD level where $\epsilon_\text{d}$ is the bare dot energy and  $\alpha= e C_\text G/(C_\text L+C_\text R+C_\text G)$ the lever arm for the gate potential $V_G$. Unprimed (primed) amplitudes correspond to particles incident from the left (right) lead.

We consider the case with a pure AC-voltage $V(t)=V\cos(\omega t)$ at contact $R$, while contact $L$ is grounded and the gate contact is kept at the constant potential $V_\text G$ corresponding to the experimental settings. The case with a pure DC-voltage bias is discussed below. As a result of the oscillating potential $V(t)$, a potential $U(t)$ is induced on the QD. The effect of the oscillating potentials is that electrons can pick up or loose quanta of energy $\hbar \omega$ when scattering at the QD. 

Our focus is on the regime of weak microwave drive, where the response is linear in the 
potentials. In this regime $V\ll \hbar \omega$ and only a single quantum can be picked up or lost. As a consequence the time dependent particle current at lead L/R has only a single Fourier component, 
\begin{equation}
I_\text{L/R}(t)=I_\text{L/R}(\omega)e^{i\omega t}+I_\text{L/R}^*(\omega)e^{-i\omega t}.
\end{equation}
The current component $I_\text{L/R}(\omega)$ can be expressed in terms of the scattering amplitudes in Eq. (\ref{smat}) and the lead Fermi distribution $f(E)$ as
\begin{equation}
I_\text L(\omega)=\frac{e^2}{h}\int dE \{-\left[1-r^*(E)r(E+\hbar \omega)\right]U(\omega) 
-t'^*(E)t'(E+\hbar \omega)\left[V(\omega)-U(\omega)\right]\}F(E,\omega),
\label{IL}
\end{equation}  
and
\begin{equation}
I_\text R(\omega)=\frac{e^2}{h}\int dE \{\left[1-r'^*(E)r'(E+\hbar \omega)\right]\left[V(\omega)-U(\omega)\right] 
+t^*(E)t(E+\hbar \omega)U(\omega)\}F(E,\omega) ,
\label{IR}
\end{equation}  
where
\begin{equation}
F(E,\omega)=\frac{f(E)-f(E+\hbar \omega)}{\hbar \omega}, \hspace{0.2cm} f(E)=\frac{1}{1+e^{E/k_\text B T}},
\end{equation}
and we have introduced the Fourier components $V(\omega)$ and $U(\omega)$ of the potentials $V(t)$ and $U(t)$. We note that $V(\omega)=V/2$, independent on $\omega$, but the frequency dependent notation is kept for convenience. 

Inserting the scattering amplitude expressions in Eq. (\ref{smat}) into the current components in Eqs. (\ref{IL}) and (\ref{IR}) we can write
\begin{equation}
I_\text L(\omega)=G(\omega)\left[\frac{i\hbar\omega}{\Gamma_{\text R}}U(\omega)-V(\omega)\right],
\end{equation}
and
\begin{equation}
I_\text R(\omega)=G(\omega)\left[\frac{i\hbar \omega}{\Gamma_{\text L}}U(\omega)+\left(1-\frac{i\hbar \omega}{\Gamma_{\text L}}\right)V(\omega)\right], 
\label{IRcurr}
\end{equation}
where
\begin{equation}
G(\omega)=\frac{e^2}{h}\int dE \mathcal{T}(E,\omega)F(E,\omega),
\label{grel}
\end{equation}
and
\begin{equation}
\mathcal{T}(E,\omega)=\frac{\Gamma_L\Gamma_R}{E-\epsilon+i(\Gamma_L+\Gamma_R)/2} 
\frac{1}{E+\hbar \omega-\epsilon-i(\Gamma_L+\Gamma_R)/2}.
\end{equation}

For non-zero frequencies, the particle currents flowing into the QD typically do not add up to zero, i.e. $I_\text L(\omega)+I_\text R(\omega) \neq 0$. As a consequence, there is nonzero charge $Q(t)$ on the QD dot which induces AC screening, or displacement, currents flowing between the QD and the leads L/R as well as the gate G. The total screening current into the QD is given by $I_\text{sc}(t)=dQ(t)/dt$, with the charge determined from classical electrostatical considerations, via the potentials $V(t)$ and $U(t)$ and the capacitances $C_\text L$, $C_\text R$, $C_\text G$. This gives the screening current Fourier component $I_\text{sc}(\omega)$ as 
\begin{equation}
I_\text{sc}(\omega)=-i\omega \left[-U(\omega)(C_\text L+C_\text R+C_\text G)+V(\omega)C_\text R\right].
\label{disp}
\end{equation}
The induced QD potential $U(\omega)$ can then be determined from the condition that the total current flowing into the dot is conserved, 
\begin{equation}
I_\text L(\omega)+I_\text R(\omega)+I_\text {sc}(\omega)=0,
\end{equation}
giving
\begin{equation}
U(\omega)=\frac{\Gamma_\text R(\hbar G(\omega)+C_R\Gamma_\text L)}{\hbar G(\omega)(\Gamma_\text L+\Gamma_\text R)+(C_\text L+C_\text R+C_\text G)\Gamma_\text R\Gamma_\text L}V(\omega).
\label{pot}
\end{equation}
We note that in the limit $\omega \rightarrow \infty$, we have $G(\omega)\rightarrow 0$ and $U(\omega)=C_\text{R}/(C_\text L+C_\text R+C_\text G) V(\omega)$, purely capacitive voltage division. The sought admittance is given by 
\begin{equation}
Y(\omega)=\frac{I_R(\omega)-i \omega C_R[V(\omega)-U(\omega)]}{V(\omega)},
\end{equation}
as the ratio of the total current, i.e., the sum of the particle and screening current, flowing into the QD from contact R and the potential at R.
 Inserting the expression of $I_R(\omega)$ in Eq. (\ref{IRcurr}) and $U(\omega)$ in (\ref{pot}) we arrive at
\begin{equation}
Y(\omega)=G(\omega)-\frac{i\omega\left[\hbar G(\omega)+\Gamma_\text L C_\text R\right]\left[\hbar G(\omega)+\Gamma_\text R(C_\text L+C_\text G)\right]}{\Gamma_\text L\Gamma_\text R(C_\text L+C_\text R+C_\text G)+\hbar G(\omega)(\Gamma_\text L+\Gamma_\text R)}.
\label{GR}
\end{equation}
This is the expression used for the numerical evaluations in the main text.

For the life-time broadening cases $\Gamma_\text{L} + \Gamma_\text{R} \gg k_\text{B} T$, effectively taking $T=0$, we can evaluate the integral in Eq. (\ref{grel}) giving
\begin{equation}
G(\omega)=\frac{e^2}{h}\frac{i\Gamma_L\Gamma_R}{\hbar \omega(\Gamma_L+\Gamma_R-i\hbar \omega)} 
 \ln \left[\frac{\epsilon^2+([\Gamma_L+\Gamma_R]/2-i\hbar \omega)^2}{\epsilon^2+(\Gamma_L+\Gamma_R)^2/4}\right].
\end{equation}
We note that at $\omega \rightarrow 0$ we have $I_\text R(0)=-I_\text L(0)=G(0)V(0)$, where  
\begin{equation}
G(0)=\frac{e^2}{h}\frac{\Gamma_L\Gamma_R}{\epsilon^2+(\Gamma_L+\Gamma_R)^2/4}
\label{DC_Lifetime_fit}
\end{equation}
is the known DC-bias conductance, as expected. 

\section{Appendix C: Sequential Tunneling Model with Periodic Voltage Drive in the Classical Limit}
\label{Appendix C}
Another extensively used model to describe QD transport builds on sequential tunneling of the electrons. Within this model, charging effects are intrinsically accounted for, while the lifetime broadening effects are neglected. This appendix calculates the QD admittance $Y(\omega)$ within the sequential tunneling approach for an applied, time periodic voltage $V(t)=V_\text{dc}+V_{ac}\sin(\omega t)$. We follow closely the work of Bruder and Sch\"oller \cite{BruderAndShoeller}, fully accounting for Coulomb blockade effects, and presenting only the main steps in the derivation. The starting point is a rate equation for the Fourier components of the probabilities $P_0(t)$ and $P_1(t)$ to have 0 or 1 extra electron on the dot. Writing the Fourier series
\begin{equation}
P_j(t)=\sum_{m=-\infty}^{\infty}\tilde P_j(m)e^{-im\omega t}, \hspace{0.5cm} j=0,1
\end{equation}
and noting that $\tilde P_j^*(m)=\tilde P_j(-m)$, the rate equation can be written as
\begin{equation}
-im\hbar \omega \tilde P_0(m)=N(\Gamma_L+\Gamma_R)\tilde P_1(m) 
-\sum_{\alpha=L,R}\sum_{n=-\infty}^{\infty} A_{nm}^{\alpha}\left[\tilde P_0(n)+N\tilde P_1(n)\right],
\label{MEdot}
\end{equation}
where
\begin{equation}
A_{nm}^{\alpha}=F_{n,m}^{\alpha}+\left(F_{-n,-m}^{\alpha}\right)^*,  
F_{nm}^{\alpha}=\frac{i^{m-n}}{2}\sum_{k=-\infty}^{\infty}J_{k+n}\left(\frac{eV_\alpha}{\hbar \omega}\right)J_{k+m}\left(\frac{eV_\alpha}{\hbar \omega}\right)f_{\alpha}(\epsilon_k), 
\end{equation}
and $J_n(x)$ is the Bessel function and $f_\alpha(\epsilon_k)$ the Fermi function of lead $\alpha=L,R$ (incorporating the dc-bias $V_\text{dc}$ at contact R) and $\epsilon_k=\epsilon+k\hbar \omega$ with $\epsilon$ the dot energy. The AC-potential drops $V_\text{L/R}$ across the L/R barriers are $V_L=-U, V_R=V_{ac}-U$ where $U=V_{ac}C_R/(C_R+C_L)$ is the amplitude of the induced dot potential. Here we have assumed $C_G \ll C_R, C_L$ as is the case for the studied devices. The integer $N$ arises from the degeneracy of the QD energy level~\cite{Hofmann2016,Barker2022}, i.e. it sets the number of possible charge states in the dot, effectively multiplying the out tunneling rates. For our spin degenerate QD we have $N = 2$. Probability conservation condition, from $P_0(t)+P_1(t)=1$ for any $t$, further gives
\begin{equation}
\tilde P_0(m)+\tilde P_1(m)=\delta_{m0}.
\end{equation}
The Fourier component of the particle current $I_\alpha(t)$ in lead $\alpha$ is given by
\begin{equation}
\tilde I_\alpha(m)=\frac{e\Gamma_\alpha}{\hbar}\left(N \tilde P_1(m)-\sum_{n=-\infty}^{\infty} A_{nm}^{\alpha}\left[\tilde P_0(n)+N\tilde P_1(n)\right]\right).
\end{equation}
As in the life-time broadened limit, the total current is obtained by the sum of the particle currents and the displacement, or screening, currents. However, as pointed out in Ref.~\citealp{BruderAndShoeller}, the screening currents are typically very small in Coulomb blockaded systems and we neglect them here. This gives the expression for the Fourier components of the total current
\begin{equation}
\tilde I(m)=\frac{C_R}{C_L+C_R}\tilde I_L(m)-\frac{C_L}{C_L+C_R}\tilde I_R(m),
\label{totcurrcomp}
\end{equation}
valid for arbitrarily drive amplitude and frequency.

Similar to the Landauer-Büttiker theory in Appendix B, we focus on the regime of small ac-driving amplitude, $eV_{ac} \ll \hbar \omega$. Expanding the Bessel functions $J_n$ in the argument $eV_{ac}/(\hbar \omega)$ and writing the probabilities $\tilde P_j(m)=\tilde P_j^{(0)}(m)+(eV_{ac}/[\hbar \omega])\tilde P_j^{(1)}(m)+... $, we can solve Eq. (\ref{MEdot}) order by order in $eV_{ac}/(\hbar \omega)$. To evaluate the DC-current $\tilde I(0)$ (in the absence of AC-drive) as well as the first AC-component $\tilde I(1)$, we need only the non-zero probability components $\tilde P_j^{(0)}(0)=1-\tilde P_1^{(0)}(0)$ (at $V_\text{ac}=0$) and $\tilde P_j^{(1)}(1)=-\tilde P_1^{(1)}(1)$ (at $V_\text{dc}=0$), given by
\begin{equation}
\tilde P_0^{(0)}(0)=\frac{N\left[\Gamma_L(1-f_L)+\Gamma_R(1-f_R)\right]}{\Gamma_Lf_L+\Gamma_Rf_R+N\left[\Gamma_L(1-f_L)+\Gamma_R(1-f_R)\right]},
\end{equation}
where we for shortness write $f_L=f_L(\epsilon), f_R=f_R(\epsilon)$, and
\begin{equation}
\tilde P_0^{(1)}(1)=\frac{-ie}{2\hbar \omega}\frac{\left[\Gamma_L V_L+\Gamma_R V_R\right]}{(\Gamma_L+\Gamma_R)\left[N+(1-N)f(\epsilon)\right]-i\hbar \omega}
 \left[N+(1-N)\tilde P_0^{(0)}(0)\right]\left[f(\epsilon_1)-f(\epsilon_{-1})\right].
\end{equation}
Within the same small amplitude approximation we have the current components
\begin{equation}
\tilde I_\alpha(0)=\frac{e\Gamma_\alpha}{\hbar}\left(N\tilde P_1^{(0)}(0)-f_{\alpha}\left[N+(1-N) \tilde P_0^{(0)}(0)\right]\right)
\end{equation}
and
\begin{equation}
\tilde I_\alpha(1)=\frac{e\Gamma_\alpha}{\hbar}\left(-\left[N+(1-N)f(\epsilon)\right]\tilde P_0^{(1)}(1) \right.  
-\left. \frac{ie V_\alpha}{2\hbar \omega}\left[N+(1-N)\tilde P_0^{(0)}(0)\right]\left[f(\epsilon_1)-f(\epsilon_{-1})\right]\right).
\end{equation}
Inserting the expressions for the probabilities and noting that $\tilde I_L(0)=-\tilde I_R(0)$, the DC-current becomes
\begin{equation}
\tilde I(0)=e\frac{N\Gamma_L\Gamma_R}{\Gamma_L+\Gamma_R}\frac{f_R-f_L}{(1-N)(f_L\Gamma_L+f_R\Gamma_R)+N(\Gamma_R+\Gamma_R)}.
\end{equation}
Expanding the Fermi distributions to first (linear) order in dc-bias voltage we arrive at the linear conductance
\begin{equation}
G=\tilde I(0)/V_{dc}=-\frac{e^2}{2\pi}\frac{N\Gamma_L\Gamma_R}{\Gamma_L+\Gamma_R}\frac{df(\epsilon)}{d\epsilon}\frac{1}{N+(1-N)f(\epsilon)}.
\label{eq:diffCondBS}
\end{equation}
which is the expression used in the plot of Fig.~\ref{fig:SymQD}~c) in the main article. 
With $N = 1$, this expression matches the standard non-degenerate result, found e.g. in Ref.~\citealp{Ihn2010}, and the Landauer-Büttiker result in the limit of $\Gamma_\text{L} + \Gamma_\text{R} \ll k_\text{B} T$ and $\omega \rightarrow 0$.

For the AC-current, we can write $\tilde I(1)=iY(\omega) V_{ac}/2$, with the admittance ($C_\Sigma=C_L+C_R$)
\begin{equation}
Y(\omega)= \frac{e^2}{2\pi} \frac{N\left[f(\epsilon_1)-f(\epsilon_{-1})\right]}{\hbar \omega \left[N+f(\epsilon)(1-N)\right]}
 \frac{\Gamma_L\Gamma_R\left[N+f(\epsilon)(1-N)\right]-i\hbar\omega\left[\Gamma_L\frac{C_R^2}{C_\Sigma^2}+\Gamma_R\frac{C_L^2}{C_\Sigma^2}\right]}{\left[(\Gamma_L+\Gamma_R)\left[N+(1-N)f(\epsilon)\right]-i\hbar\omega \right]},
\label{eq:BnS_ST_theory_equation}
\end{equation}
which is the expression used for the fit of Fig.~\ref{fig:SymQD}~d) in the main article. We note that for $\omega \rightarrow 0$ we have $Y(0)=G$, as expected.





\section{Appendix D: Sequential Tunneling Model with P(E) Theory}
\label{Appendix D}
The sequential tunneling model of Appendix C treats the voltage drive $V(t)$ as a purely classical signal. Here we consider briefly another sequential tunneling model approach, the $P(E)$ theory, that treats the voltage in the resonator quantum mechanically.
To calculate the electron and photon transport properties of the studied system within the $P(E)$ theory, we follow the formalism of Refs.~\citealp{Ingold1992,maisi2014b,Souquet2014}. In the presence of a photon environment, the tunneling rate $\Gamma_+$ into the quantum dot (QD) from an electronic reservoir and the opposite rate $\Gamma_-$ out from it~\cite{Hofmann2016,Barker2022} are convoluted with the probability $P(E)$ to absorb energy $E$ from the environment. The resulting tunneling rates for left, $i = L$, and right, $i = R$, tunnel junction are
\begin{equation}
\label{eq:GammaNQD}
\left\{
\begin{array}{ccl}
\Gamma_{i+}(\varepsilon_i) &=& \displaystyle \frac{\Gamma_i}{h} \int_{-\infty}^\infty dE\; P(E) f(\varepsilon_i + E) \vspace{3pt} \\
\Gamma_{i-}(\varepsilon_i) &=& \displaystyle N\; \frac{\Gamma_i}{h} \int_{-\infty}^\infty dE\; P(-E) \left[ 1-f(\varepsilon_i + E) \right].
\end{array}
\right.
\end{equation}
Here $\Gamma_i$ is the tunnel coupling strength, $\varepsilon_i$ the QD energy level position with respect to the reservoir Fermi level, $f(E)$ is the Fermi function defining the electron occupation distribution in the reservoir, and the additional pre-factor $N$ for tunneling out arises from the degeneracy of the considered QD energy level~\cite{Hofmann2016,Barker2022}, as above. For our system, we have $N=2$ as there are two electrons to choose from to tunnel out and only one vacant state to tunnel into the QD.

The $P(E)$ function is set by the environment. In our case, the resonator forms a single photon mode at frequency $\omega_r$ as the environment~\cite{Girvin2014,Souquet2014}. For the characteristic impedance $Z_0 = 53~\Omega$ of our resonator, we have $z = \pi Z_0 G_0 \ll 1$, where $G_0 = e^2/h$ is the conductance quantum. For the coherent drive used in the experiment with low enough input power and at low temperature $kT \ll \hbar \omega_r$, the resonator is in a coherent state with small photon number $\left<n\right> \ll 1/z$. In this limit, the $P(E)$ function reads~\cite{Souquet2014}
\begin{equation}
\label{eq:PEsinglephoton}
P(E) = \Big[ 1 - z \nu^2 \big(2 \left<n\right> + 1 \big)  \Big] \delta(E) + z \nu^2 \left<n\right>\, \delta(E + \hbar\omega) + z \nu^2 \big(\left<n\right> + 1 \big) \delta(E - \hbar\omega),
\end{equation}
where $\nu$ is the fraction of the resonator voltage that appears accross the tunnel barrier where the tunneling takes place~\cite{Childress2004}. We assume that the junction capacitances of the quantum dot are equal. In addition, the gate capacitance is much smaller than the junction capacitances in the studied devices. Under these conditions, the resonator voltage is divided evenly over the two tunnel barriers, and $\nu = 1/2$ and the $P(E)$ function is the same for the two tunnel junctions.

The three terms in Eq.~(\ref{eq:PEsinglephoton}) correspond to 1) no interactions with the photons, 2) absorption of a photon and 3) emission of a photon with probabilities $P_0 = 1 - z \nu^2 \left(2 \left<n\right> + 1\right)$, $P_{-1} = z \nu^2 \, \left<n\right>$ and $P_{+1} = z \nu^2 \left(\left<n\right> + 1 \right)$ respectively. Interestingly, the photon absorption or emission probability is solely set by the impedance $z$, coupling $\nu$ and photon number $\left<n\right>$ of the microwave resonator and does not depend on the tunnel coupling $\Gamma_i$. The absorption and emission rates, however, depend on $\Gamma_i$ via Eq.~(\ref{eq:GammaNQD}). Note also that Eq.~(\ref{eq:PEsinglephoton}) contains only single photon absorption and emission processes. Multiphoton processes are suppressed in the low drive limit $\left<n\right> \ll 1/(z \nu^2)$, in line with the experiments of Ref.~\citealp{Haldar2022}.

\section{Tunneling rates and electrical conduction}
With Eqs.~(\ref{eq:GammaNQD}-\ref{eq:PEsinglephoton}), we obtain the tunneling rates as
\begin{equation}
\label{eq:GammaP}
\left\{
\begin{array}{ccl}
\Gamma_{i+}(\varepsilon_i) &=&  \displaystyle \frac{\Gamma_i}{h}  \Big[ P_0 f(\varepsilon_i) + P_{-1} f(\varepsilon_i-\hbar \omega) +  P_{+1} f(\varepsilon_i+\hbar \omega)  \Big] \vspace{3pt} \\
\Gamma_{i-}(\varepsilon_i) &=& N  \displaystyle \frac{\Gamma_i}{h}  \Big[ P_0 [1-f(\varepsilon_i)] + P_{-1} [1-f(\varepsilon_i+\hbar \omega)] +  P_{+1} [1-f(\varepsilon_i-\hbar \omega)]  \Big] = N \Gamma_{i+}(-\varepsilon_i).
\end{array}
\right.
\end{equation}
 The energy differences across the two junctions are $\varepsilon_L = \varepsilon_d + \alpha V_G - eV_\mathrm{b}/2$ and $\varepsilon_R = \varepsilon_d + \alpha V_G + eV_\mathrm{b}/2$, where $\alpha$ is the gate lever arm, $V_g$ the voltage applied to a gate electrode and $V_b$ the bias voltage between the source and drain. Here we have again assumed that the junction capacitances are equal and the gate capacitance is small compared to the junction capacitances which divide the bias voltage $V_b$ evenly over the junctions.

Next we determine the electrical conductance $G$ by setting a rate equation to describe the probabilitiy $p$ to have an excess electron in the QD and $(1-p)$ of not having the excess electron. This rate equation reads
\begin{equation}
\label{eq:rateEq}
\frac{dp}{dt} = -\Gamma_-\, p + \Gamma_+\, (1-p),
\end{equation}
where $\Gamma_- = \Gamma_\mathrm{L-}(\varepsilon_\mathrm{L}) + \Gamma_\mathrm{R-}(\varepsilon_\mathrm{R})$ is the sum of the rates for the electron to tunnel out and  $\Gamma_+ = \Gamma_\mathrm{L+}(\varepsilon_\mathrm{L}) + \Gamma_\mathrm{R+}(\varepsilon_\mathrm{R})$ the sum of the rates to tunnel in. The steady-state solution, $dp/dt = 0$, is
\begin{equation}
\label{eq:p}
p = \frac{\Gamma_+}{\Gamma_+ + \Gamma_-}.
\end{equation}
The electrical current $I$ through the QD is given by
\begin{equation}
\label{eq:p}
I = e \Big[ \Gamma_\mathrm{L+}(\varepsilon_\mathrm{L})\, (1-p) - \Gamma_\mathrm{L-}(\varepsilon_\mathrm{L})\, p \Big] = e
\frac{\Gamma_\mathrm{L+}(\varepsilon_\mathrm{L})\Gamma_\mathrm{R-}(\varepsilon_\mathrm{R}) - 
\Gamma_\mathrm{L-}(\varepsilon_\mathrm{L}) \Gamma_\mathrm{R+}(\varepsilon_\mathrm{R})}
{\Gamma_\mathrm{L+}(\varepsilon_\mathrm{L}) + \Gamma_\mathrm{R+}(\varepsilon_\mathrm{R}) + 
\Gamma_\mathrm{L-}(\varepsilon_\mathrm{L}) + \Gamma_\mathrm{R-}(\varepsilon_\mathrm{R})}. 
\end{equation}
Finally, the differential conductance is obtained as
\begin{equation}
\label{eq:p}
G = \frac{dI}{dV_\mathrm{SD}}.
\end{equation}
For small bias voltage, $eV_\mathrm{SD} \ll kT$, and low photon absorption and emission probabilities, $P_{-1}, P_{+1} \ll P_0$, we obtain
\begin{equation}
\label{eq:p}
G = \frac{e^2}{hkT}\; \frac{\Gamma_\mathrm{L}\Gamma_\mathrm{R}}{\Gamma_\mathrm{L} + \Gamma_\mathrm{R}}\; \frac{1}{1 + \frac{1}{N} + \frac{2}{\sqrt{N}} \cosh\left((\varepsilon_d + \alpha V_G)/kT +   \frac{1}{2} \ln N \right)},
\end{equation}
which is the zero bias conductance $G$ for a QD with degeneracy $N$, and matches Eq.~(\ref{eq:diffCondBS}).
 Compared to the standard non-degenerate result with $N = 1$, the degeneracy $N \neq 1$ keeps the lineshape the same but displaces the peak position by $(kT \ln N)/2$ in energy and increases the maximum value by a factor of $4/(1+ 1/\sqrt{N})^2$. For $N = 2$, these are both small effects: $(kT \ln N)/2 \approx 0.35\; kT$ and $4/(1+ 1/\sqrt{N})^2 \approx 1.37$. With photon absorption accounted for, two side peaks appear in $G$. These are separated by $\hbar \omega$ in energy from the main conduction peak. The size of these ones are $P_{-1}/P_0$ relative to the main peak, and can therefore be neglected for our devices with $P_{-1} \ll P_0$.

\section{Photon absorption and emission rates}
In the above, we determined the low frequency electrical conductance $G$. By collecting the photon absorption terms from Eq.~(\ref{eq:GammaP}), we obtain the photon absorption rate as
\begin{equation}
\label{eq:gammaPhoton}
\Gamma_-^\mathrm{photon} = P_{-1} \Big[ \big( \Gamma_\mathrm{L}\; f(\varepsilon_L-\hbar \omega) + \Gamma_\mathrm{R}\; f(\varepsilon_R-\hbar \omega) \big) (1-p) + N \big( \Gamma_\mathrm{L}\; f(-\varepsilon_L-\hbar \omega) + \Gamma_\mathrm{R}\; f(-\varepsilon_R-\hbar \omega) \big) p \Big]/h,
\end{equation}
where we have summed the absorption terms with the corresponding weights of $(1-p)$ and $p$ of the probability to be in the right starting state of the QD. The absorption rate per photon corresponds to a resonator loss term $\kappa_\mathrm{QD}$ which is thus
\begin{equation}
\label{eq:gammaPhoton}
\kappa_\mathrm{QD} = \frac{\Gamma_-^\mathrm{photon}}{\left<n\right>} = \nu^2 z \Big[ \big( \Gamma_\mathrm{L}\; f(\varepsilon_L-\hbar \omega) + \Gamma_\mathrm{R}\; f(\varepsilon_R-\hbar \omega) \big) (1-p) + N \big( \Gamma_\mathrm{L}\; f(-\varepsilon_L-\hbar \omega) + \Gamma_\mathrm{R}\; f(-\varepsilon_R-\hbar \omega) \big) p \Big]/h.
\end{equation}
For zero bias voltage, $V_\mathrm{SD} = 0$, at low temperature, $kT \ll \hbar \omega$, and small photon number $\left<n\right>$, we have $\varepsilon_L = \varepsilon_R$ and the photon absorption takes place in two distinct regimes. The first one is with $-\hbar \omega < \varepsilon_i < 0$. Here we have $p = 1$, i.e. the quantum dot is occupied essentially at all times and $f(-\varepsilon_{i}-\hbar \omega) = 1$. With this, Eq.~(\ref{eq:gammaPhoton}) yields
\begin{equation}
\label{eq:kappan}
\kappa_\mathrm{QD} = z \nu^2 (\Gamma_\mathrm{L}+\Gamma_\mathrm{R})N/h.
\end{equation}
The other regime takes place for $0 < \varepsilon_i < \hbar \omega$. In this case we have the QD essentially always unoccupied, i.e. $p = 0$ while the energy $\varepsilon_i$ is still low enough so that $f(\varepsilon_{i}-\hbar \omega) = 1$. Now we have 
\begin{equation}
\label{eq:kappap}
\kappa_\mathrm{QD} = z \nu^2 (\Gamma_\mathrm{L}+\Gamma_\mathrm{R})/h,
\end{equation}
which is the same as above but smaller by the degeneracy factor $N$. The sum $(\Gamma_\mathrm{L}+\Gamma_\mathrm{R})$ reflects the fact that with no bias voltage applied, the two junctions are effectively in parallel. The dashed lines of Fig.~\ref{fig:SymQD} d) of the main article are the results of Eqs.~(\ref{eq:kappan}) and (\ref{eq:kappap}) and show that the measured response follows these values. The fitted full curve in the figure, based on Appendix C, is essentially the same as the result of Eq.~(\ref{eq:gammaPhoton}). This demonstrates that the two sequential tunneling models yield identical results in this regime. For a stronger coupling via larger impedance $Z_0$, deviances are expected: in the absence of the drive, the calculation of Appendix C predicts still no changes to the standard dc transport result while the $P(E)$ theory has a strong spontaneous emission process and thus a suppressed probability $P_0$ for the dc transport known as the dynamical Coulomb blockade~\cite{Altimiras2014}.




\end{document}